\documentclass[11pt]{article}

\usepackage{amsmath}

\newcommand{\N}{N\raise.7ex\hbox{\underline{$\circ $}}$\;$}

\begin{document}

\begin{center}
{\bf A.A. Bogush, V.M. Red'kov, N.G. Tokarevskaya, George J.
Spix\\
Matrix-based approach to electrodynamics in media}
\\
 B.I. Stepanov Institute of Physics \\
National Academy of Sciences of Belarus  \\
BSEE Illinois Institute of Technology, USA\\
redkov@dragon.bas-net.by , gjspix@msn.com

\end{center}

\begin{quotation}

The Riemann -- Silberstein -- Majorana -- Oppenheimer approach to
the Maxwell electrodynamics in presence of electrical sources and
arbitrary media is investigated within the  matrix formalism. The
symmetry of the matrix Maxwell equation under transformations of
the complex rotation group SO(3.C) is demonstrated explicitly. In
vacuum case, the matrix form includes  four  real $4 \times 4$
matrices $\alpha^{b}$. In presence of media matrix form requires
two  sets of   $4 \times 4$ matrices, $\alpha^{b}$ and $\beta^{b}$
-- simple and symmetrical realization of which is given. Relation
of  $\alpha^{b}$ and $\beta^{b}$  to   the Dirac matrices in
spinor basis is found. Minkowski constitutive relations in case of
any linear media are given in a short algebraic
 form based on the use  of complex 3-vector fields and
complex orthogonal rotations from SO(3.C) group. The  matrix
complex formulation in  the Esposito's form,based on the use of
two electromagnetic 4-vectors, $e^{\alpha}(x) = u_{\beta}
F^{\alpha \beta}(x) ,  b^{\alpha} (x) = u_{\beta}
\tilde{F}^{\alpha \beta}(x) $ is studied and discussed. It is
argued that Esposito form is achieved  trough  the use of a
trivial identity $I=U^{-1}(u)U(u)$ in the Maxwell equation.

\end{quotation}

Pacs: 02.20.-a, 03.50.De


 rotation group, Lorentz group, constitutive relations

\section{Introduction}

Special relativity arose from
study of the symmetry properties of the Maxwell equations with respect to
motion of references frames: Lorentz  \cite{1904-Lorentz},  Poincar'e \cite{1905-Poincare},
Einstein \cite{1905-Einstein}
Naturally, an analysis of the Maxwell equations with respect to Lorentz transformations was the first objects
of relativity theory:
 Minkowski \cite{1908-Minkowski},
 Silberstein \cite{1907-Silberstein(1)}-\cite{1907-Silberstein(2)},
Marcolongo \cite{1914-Marcolongo},  Bateman  \cite{1915-Bateman}, and  Lanczos \cite{1919-Lanczos},
Gordon \cite{1923-Gordon}, Mandel'stam --  Tamm \cite{1925-Mandel'stam}-\cite{1925-Tamm(1)}-\cite{1925-Tamm(2)}.

After discovering the relativistic equation for a particle with
spin 1/2 -- Dirac \cite{1928-Dirac} -- much work was done to study
spinor and vectors within the Lorentz group theory: M\"{o}glich
\cite{1928-Moglich},  Ivanenko -- Landau
\cite{1928-Ivanenko-Landau}, Neumann \cite{1929-Neumann}, van der
Waerden \cite{1929-Waerden}, Juvet \cite{1930-Juvet}. As was shown,
any quantity which transforms linearly under Lorentz
transformations is a spinor. For that reason spinor quantities are
considered as fundamental in quantum field theory and basic
equations for such quantities should be written in a spinor form.
A spinor formulation of Maxwell equations was studied by Laporte
and Uhlenbeck \cite{1931-Laporte}, also see Rumer
\cite{1936-Rumer}. In 1931,   Majorana \cite{1931-Majorana} and
Oppenheimer \cite{1931-Oppenheimer} proposed to consider the
Maxwell theory of electromagnetism as the wave mechanics of the
photon. They introduced a  complex 3-vector wave function
satisfying the massless Dirac-like equations. Before Majorana  and
Oppenheimer,    the most crucial steps were made by Silberstein
\cite{1907-Silberstein(1)}, he showed the possibility to have
formulated Maxwell equation in terms of complex 3-vector entities.
Silberstein in his second paper  \cite{1907-Silberstein(2)} writes
that the complex form of Maxwell equations has been known before;
he refers there to the second volume of the lecture notes on the
differential equations of mathematical physics by B. Riemann that
were edited and published by H. Weber in 1901 \cite{1901-Weber}.
 This not widely used fact  is  noted by Bialynicki-Birula   \cite{1994-Bialynicki-Birula}).

Maxwell equations in the  matrix Dirac-like  form considered during long time by many authors, the
interest to the Majorana-Oppenheimer formulation of electrodynamics has grown in recent
years:

\begin{quotation}

Luis de Broglie \cite{1934-Broglie(1)}-\cite{1934-Broglie(2)}-\cite{1939-Broglie}-\cite{1940-Broglie},
 Petiau \cite{1936-Petiau}, Proca \cite{1936-Proca}- \cite{1946-Proca},
Duffin \cite{1938-Duffin}, Kemmer \cite{1939-Kemmer}-\cite{1943-Kemmer}-\cite{1960-Kemmer},  Bhabha \cite{1939-Bhabha},
Belinfante \cite{1939-Belinfante(1)}-\cite{1939-Belinfante(2)},
Taub \cite{1939-Taub},  Sakata  -- Taketani \cite{1940-Sakata}, Schr\"{o}dinger \cite{1943-Schrodinger(1)}-\cite{1943-Schrodinger(2)},
Heitler \cite{1943-Heitler},  \cite{1946-Harish-Chandra(1)}-\cite{1946-Harish-Chandra(2)},
Mercier \cite{1949-Mercier}, Imaeda \cite{1950-Imaeda},   Fujiwara \cite{1953-Fujiwara}, Ohmura  \cite{1956-Ohmura},
  Borgardt \cite{1956-Borgardt}-\cite{1958-Borgardt},  Fedorov \cite{1957-Fedorov},  Kuohsien \cite{1957-Kuohsien},
Bludman   \cite{1957-Bludman}, Good \cite{1957-Good},  Moses \cite{1958-Moses}-\cite{1959-Moses}-\cite{1973-Moses},
Lomont \cite{1958-Lomont},
Bogush -- Fedorov \cite{1962-Bogush-Fedorov}, Sachs -- Schwebel  \cite{1962-Sachs-Schwebel},   Ellis \cite{1964-Ellis},
Oliver \cite{1968-Oliver},
Beckers -- Pirotte \cite{1968-Beckers},  Casanova \cite{1969-Casanova},  Carmeli \cite{1969-Carmeli},
Bogush \cite{1971-Bogush}, Lord \cite{1972-Lord},
Weingarten \cite{1973-Weingarten},   Mignani --  Recami --  Baldo  \cite{1974-Recami1},
Newman \cite{1973-Newman}, \cite{1974-Frankel},  \cite{1975-Jackson}, Edmonds \cite{1975-Edmonds},
Silveira \cite{1980-Silveira},   Jena --   Naik --  Pradhan \cite{1980-Jena},  Venuri  \cite{1981-Venuri},
Chow \cite{1981-Chow}, Fushchich -- Nikitin  \cite{1983-Fushchich}, Cook \cite{1982-Cook(1)}-\cite{1982-Cook(2)},
Giannetto \cite{1985-Giannetto},   -- Y\'epez,
Brito -- Vargas  \cite{1988-Nunez}, Kidd --  Ardini --  Anton  \cite{1989-Kidd}, Recami \cite{1990-Recami},
Krivsky --  Simulik \cite{1992-Krivsky}, Inagaki \cite{1994-Inagaki},
Bialynicki-Birula \cite{1994-Bialynicki-Birula}-\cite{1996-Bialynicki-Birula}-\cite{2005-Birula},
Sipe \cite{1995-Sipe}, \cite{1996-Ghose}, Esposito \cite{1998-Esposito},
Dvoeglazov \cite{1998-Dvoeglazov} (see a big list of  relevant references therein)-\cite{2001-Dvoeglazov},
Gersten  \cite{1998-Gersten},
Kanatchikov \cite{2000-Kanatchikov}, Gsponer \cite{2002-Gsponer},
\cite{2001-Ivezic(1)}-\cite{2002-Ivezic}-\cite{2002-Ivezic(2)}-\cite{2002-Ivezic(3)}-\cite{2003-Ivezic}-
\cite{2005-Ivezic(1)}-\cite{2005-Ivezic(2)}-\cite{2005-Ivezic(3)}-\cite{2006-Ivezic}, Donev
--  Tashkova. \cite{2004-Donev(1)}-\cite{2004-Donev(2)}-\cite{2004-Donev(3)}.

\end{quotation}

Our treatment will be with a quite definite  accent:
 the main attention is given to technical aspect of classical electrodynamics
 based on the theory of rotation complex group SO(3.C) (isomorphic to the Lorentz group -- see
 Kur\c{s}uno$\hat{g}$lu \cite{1961-Kursunoglu},
Macfarlane  cite\cite{1962-Macfarlane}-\cite{1966-Macfarlane}, Fedorov \cite{1979-Fedorov}).

\section{  Complex matrix form of Maxwell theory in vacuum }

Let us start with Maxwell equations in  a uniform ($\epsilon,
\mu$)-media  in  presence  of  external sources \cite{1941-Stratton}-\cite{1961-Panofsky-Phillips}-\cite{1975-Jackson}:
\begin{eqnarray}
(F^{ab}) \qquad \mbox{div} \; c{\bf B} = 0 \; , \qquad \mbox{rot}
\;{\bf E} = -{\partial c {\bf B} \over \partial ct} \; ,
\nonumber
\\
(H^{ab}) \qquad  \mbox{div}\; {\bf E} = {\rho \over \epsilon
\epsilon_{0}} , \qquad
 \mbox{rot} \; c{\bf B} = \mu \mu_{0}c {\bf J} + \epsilon \mu
   {\partial {\bf E} \over \partial ct} \; .
\label{1.1a}
\end{eqnarray}

\noindent With the use of usual notation for  current 4-vector
$
j^{a} = (\rho, {\bf J} /c) \; ,  \; c^{2} = 1 /  \epsilon_{0}\mu_{0} \; ,
$  eqs. (\ref{1.1a}) read (first, consider the  vacuum case):
\begin{eqnarray}
 \mbox{div} \; c{\bf B} = 0 \; , \qquad \mbox{rot}
\;{\bf E} = -{\partial c {\bf B} \over \partial ct} \; ,
\nonumber
\\
 \mbox{div}\; {\bf E} = {\rho \over  \epsilon_{0}} , \qquad
 \mbox{rot} \; c{\bf B} =  {{\bf j} \over \epsilon_{0}}  +
   {\partial {\bf E} \over \partial ct} \; .
\label{1.2a}
\end{eqnarray}

Let us introduce 3-dimensional complex vector
\begin{eqnarray}
\psi^{k} =   E^{k} + i c B^{k} \; ,
\label{1.3a}
\end{eqnarray}

\noindent with the help of which the above equations can be  combined into (see  Silberstein \cite{1907-Silberstein(1)}-\cite{1907-Silberstein(2)},
 Bateman  \cite{1915-Bateman}, Majorana \cite{1931-Majorana},  Oppenheimer \cite{1931-Oppenheimer},
 and many others)
\begin{eqnarray}
\partial_{1}\Psi ^{1} + \partial_{2}\Psi ^{0} + \partial_{3}\Psi ^{3} =
j^{0} / \epsilon_{0}  \; ,\qquad
-i\partial_{0} \psi^{1} + (\partial_{2}\psi^{3} -
\partial_{3}\psi^{2}) = i\; j^{1} / \epsilon_{0} \; ,
\nonumber
\\
-i\partial_{0} \psi^{2} + (\partial_{3}\psi^{1} -
\partial_{1}\psi^{3}) = i\; j^{2} / \epsilon_{0} \; , \qquad
-i\partial_{0} \psi^{3} + (\partial_{1}\psi^{2} -
\partial_{2}\psi^{1}) =  i\; j^{3} / \epsilon_{0} \; ,
\label{1.3b}
\end{eqnarray}

\noindent  where  $x_{0}=ct, \; \partial_{0} = c \;\partial_{t}$.
These four relations can be rewritten in a  matrix form
using a 4-dimensional column  $\psi$ with one additional zero-element (Fuschich -- Nikitin \cite{1983-Fushchich}:
\begin{eqnarray}
(-i\alpha^{0} \partial_{0} + \alpha^{j} \partial_{j} ) \Psi = J \;
, \qquad \Psi = \left | \begin{array}{c} 0 \\\psi^{1} \\\psi^{2}
\\ \psi^{3}
\end{array} \right | \; , \qquad
\alpha^{0} = \left | \begin{array}{rrrr}
a_{0} & 0  &  0  & 0  \\
a_{1} & 1  &  0  & 0  \\
a_{2} & 0  &  1  & 0  \\
a_{3} & 0  &  0  & 1
\end{array}  \right | \; ,
\nonumber
\\
\alpha^{1} = \left | \begin{array}{rrrr}
b_{0} & 1  &  0  & 0  \\
b_{1} & 0  &  0  & 0  \\
b_{2} & 0  &  0  & -1  \\
b_{3} & 0  &  1  & 0
\end{array}  \right |\;  ,
\alpha^{2} = \left | \begin{array}{rrrr}
c_{0} & 0  &  1  & 0  \\
c_{1} & 0  &  0  & 1  \\
c_{2} & 0  &  0  & 0  \\
c_{3} & -1  & 0  & 0
\end{array}  \right |\;  ,
\alpha^{3} = \left | \begin{array}{rrrr}
d_{0} & 0  &  0  & 1  \\
d_{1} & 0  & -1  & 0  \\
d_{2} & 1  &  0  & 0  \\
d_{3} & 0  &  0  & 0
\end{array}  \right | \; .
\label{1.4}
\end{eqnarray}

\noindent
Here, there arise four  matrices, in which  numerical parameters  $a_{k}, b_{k}, c_{k}, d_{k}$
may be arbitrary. Our choice for the matrix form of eight  Maxwell equations  is the following:
\begin{eqnarray}
(-i \partial_{0} + \alpha^{j} \partial_{j} ) \Psi =J \; , \qquad
\Psi = \left | \begin{array}{c} 0 \\\psi^{1} \\\psi^{2} \\
\psi^{3}
\end{array} \right | \; , \qquad J=
{1 \over \epsilon_{0}} \; \left | \begin{array}{c} j^{0} \\ i\;
j^{1} \\ i\; j^{2} \\ i \; j^{3}
\end{array} \right | \; ,
\nonumber
\\
\alpha^{1} = \left | \begin{array}{rrrr}
0 & 1  &  0  & 0  \\
-1 & 0  &  0  & 0  \\
0 & 0  &  0  & -1  \\
0 & 0  &  1  & 0
\end{array}  \right |, \qquad
\alpha^{2} = \left | \begin{array}{rrrr}
0 & 0  &  1  & 0  \\
0 & 0  &  0  & 1  \\
-1 & 0  &  0  & 0  \\
0 & -1  & 0  & 0
\end{array}  \right |, \qquad
\alpha^{3} = \left | \begin{array}{rrrr}
0 & 0  &  0  & 1  \\
0 & 0  & -1  & 0  \\
0 & 1  &  0  & 0  \\
-1 & 0  &  0  & 0
\end{array}  \right |,
\nonumber
\\
(\alpha^{1})^{2} = -I \; , \qquad  (\alpha^{2})^{2} = -I \; ,  \qquad
(\alpha^{3})^{2} = -I \;  ,
\nonumber
\\
\alpha^{1} \alpha^{2}= - \alpha^{2} \alpha^{1} =  \alpha^{3} \;,
\qquad \alpha^{2} \alpha^{3} = - \alpha^{3} \alpha^{2} =
\alpha^{1}\;, \qquad \alpha^{3} \alpha^{1} = - \alpha^{1}
\alpha^{3} = \alpha^{2}\;.
\label{1.10}
\end{eqnarray}

Now let us  consider the problem of relativistic invariance of this  equation. The lack of manifest invariance of
3-vector complex form of Maxwell theory has been intensively discussed  in various aspects:
for instance, see  Esposito \cite{1998-Esposito}, Ivezic
\cite{2001-Ivezic(1)}-\cite{2002-Ivezic}-\cite{2002-Ivezic(2)}-\cite{2002-Ivezic(3)}-\cite{2003-Ivezic}-
\cite{2005-Ivezic(1)}-\cite{2005-Ivezic(2)}-\cite{2005-Ivezic(3)}-\cite{2006-Ivezic}).

Arbitrary  Lorentz transformation over the function $\Psi$ is given by (take notice that one may
introduce four undefined parameters $s_{0}, ...,s_{3}$, but we will take $s_{0}=1, s_{j}=0$ )
\begin{eqnarray}
S = \left | \begin{array}{cccc}
s_{0} & 0 & 0 & 0  \\
s_{1} & .  & . & . \\
s_{2} & .   &O(k) & . \\
s_{3} & . & . & .
\end{array} \right | \;, \qquad \Psi' = S \Psi \; , \qquad  \Psi = S^{-1} \Psi '\; ,
\label{1.13a}
\end{eqnarray}

\noindent where $O(k)$ stands for a $(3\times 3$-rotation complex matrix from
$SO(3,C)$, isomorphic to the Lorentz  group -- more  detail see in \cite{1979-Fedorov} and below in the present text.
Equation for a primed function  $\Psi'$  is
\begin{eqnarray}
(-i \partial_{0} + S \alpha^{j}S^{-1}  \partial_{j} ) \; \Psi '= S\;J \; .
\nonumber
\end{eqnarray}

\noindent
When working with matrices $\alpha^{j}$ we will use vectors
 ${\bf e}_{i}$ and ($ 3 \times 3$)-matrices $\tau_{i}$, then the structure $S \alpha^{j}S^{-1} $ is
\begin{eqnarray}
S \alpha^{j}S^{-1} =  \left | \begin{array}{cc}
0 & {\bf e}_{j} O^{-1}(k)   \\
- O(k){\bf e}^{t}_{j} &  O(k) \tau_{j} O^{-1}(k)
\end{array} \right |  = \alpha^{m} O_{mj}(k) \; .
\label{1.13c}
\end{eqnarray}

\noindent
Therefore,  the  Maxwell equation gives
\begin{eqnarray}
 (-i \partial_{0} + \alpha^{m}\;  \partial'_{m} ) \Psi'  = S J \;,
 \qquad
 O_{mj}   \partial_{j} = \partial_{m}' \;  .
\label{1.13d}
\end{eqnarray}

Now, one  should give  special attention   to the following: the  symmetry properties  given by  (\ref{1.13d})
look satisfactory only at real values of  parameter $a$ -- in this case it describes
symmetry of the  Maxwell equations under Euclidean rotations. However, if the values of $a$ are imaginary
  the above transformation  $S$  gives  a  Lorentzian boost; for instance, in the plane $0-3$ the boost is
  \begin{eqnarray}
a = i b  \; ,  \qquad S(a=ib) = \left | \begin{array}{cccc}
1  &       0   &      0  &  0  \\
0  &  ch \; b   &  -i sh\; b   &  0  \\
0  & i sh\; b    &  ch\; b  &  0  \\
0 & 0  & 0  &  1
\end{array} \right |
\label{1.15a}
\end{eqnarray}

\noindent and the formulas (\ref{1.13c}) will take the form
\begin{eqnarray}
S\alpha^{1}S^{-1} = ch\; b \; \alpha^{1} + i sh\; b  \; \alpha^{2}
\; ,
\nonumber
\\
S\alpha^{2}S^{-1} = -i sh\; b\; \alpha^{1} + ch\; b \; \alpha^{2}
\; , \qquad S\alpha^{3}S^{-1} = \alpha^{3} \;.
\label{1.15b}
\end{eqnarray}

\noindent Correspondingly, the Maxwell matrix equation after transformation (\ref{1.15a})-(\ref{1.15b})
will look  asymmetric
\begin{eqnarray}
[\; (-i \partial_{0} + \alpha^{3}  \partial_{3} ) +
 (ch\; b \; \alpha^{1} + i sh\; b  \; \alpha^{2})  \;\partial_{2} +
\nonumber
\\
 +
(-i sh\; b\; \alpha^{1} + ch\; b \; \alpha^{2} )  \; \partial_{3}
\; ] \;  \Psi'  = SJ \; .
\label{1.15c}
\end{eqnarray}

\noindent However, one  can note  an identity
\begin{eqnarray}
(ch\; b - i sh\; b \; \alpha^{3}) (-i \partial_{0} + \alpha^{3}
\partial_{3}) =
\nonumber
\\
= -i (ch\; b \; \partial _{0}  - sh\; b \; \partial_{3}) +
\alpha^{3} ( -sh\; b \; \partial_{0} + ch\; b \; \partial _{3}) =
-i \partial_{0}' + \alpha^{3} \partial_{3}' \; ,
\label{1.16a}
\end{eqnarray}

\noindent where derivatives are changed in accordance with the Lorentzian boost rule:
\begin{eqnarray}
ch\; b \; \partial _{0}  - sh\; b \; \partial_{3} =
\partial_{0}'  \; , \qquad
 -sh\; b \; \partial_{0} + ch\; b \; \partial _{3} =
\partial_{3}' \; .
\nonumber
\end{eqnarray}

\noindent Therefore. it remains to determine  the action of the  operator (we introduce special notation for it, $\Delta$)
\begin{eqnarray}
\Delta = ch\; b - i sh\; b \; \alpha^{3}
\label{1.16b}
\end{eqnarray}

\noindent on two  other terms in eq. (\ref{1.15c}) -- one might expect two relations:
\begin{eqnarray}
(ch\; b - i sh\; b \; \alpha^{3}) (ch\; b \; \alpha^{1} + i sh\;
b  \; \alpha^{2}) = \alpha^{2} \; ,
\nonumber
\\
(ch\; b - i sh\; b \; \alpha^{3}) (-i sh\; b\; \alpha^{1} + ch\; b
\; \alpha^{2} ) = \alpha^{3}  \; .
\label{1.16c}
\end{eqnarray}

\noindent As easily  verified they hold indeed.
We should   calculate the term
\begin{eqnarray}
\Delta  S \; J = \left | \begin{array}{c}
ch\; b \; j^{0} + sh\; b \; j^{3} \\
i \;j^{1} \\
i\;j^{2}\\
i( sh\; b \; j^{0} + ch\;b \; j^{3})
\end{array} \right | \; ;
\label{1.16d}
\end{eqnarray}

\noindent the right-hand side of (\ref{1.16d})  is what we need.

Thus, the symmetry of the matrix Maxwell equation under
the Lorentzian  boost in the plane $0-3$  is described by relations:
\begin{eqnarray}
\Delta (b)\; (-i \partial_{0} + S\alpha^{j}S^{-1}  \partial_{j})\;
\Psi'  = \Delta S \; J \equiv J' \; , \qquad  (-i
\partial_{0}' + \alpha^{j}  \partial_{j}') \Psi'  = J' \; ,
\nonumber
\\
 S(b) = \left | \begin{array}{cccc}
1  &       0   &      0  &  0  \\
0  &  ch \; b   &  -i sh\; b   &  0  \\
0  & i sh\; b    &  ch\; b  &  0  \\
0 & 0  & 0  &  1
\end{array} \right | \; , \qquad \Delta (b) = ch\; b - i sh\; b \; \alpha^{3} \; .
\label{1.17a}
\end{eqnarray}

For the  general case, one can think that for an arbitrary oriented boost
the  operator $\Delta$ should be of the form:
\begin{eqnarray}
\Delta  = \Delta_{\alpha} = ch\; b - i \; sh\; b \; n_{j} \;
\alpha^{j} \; .
\nonumber
\end{eqnarray}

To  verify this, one should obtain  mathematical description of that general  boost.
We  will start with the known parametrization of the real 3-dimension group \cite{1979-Fedorov})
\begin{eqnarray}
O (c)  =  I + 2 \; [ \; c_{0} \;\vec{c}\; ^{\times} + ( \vec{c}\;
^{\times})^{2} \; ]\;  , \qquad  ( \vec{c}\; ^{\times})_{kl}=
-\epsilon _{klj} \;a _{j} \; ,
\nonumber
\\
O(c)  =\left | \begin{array}{lll}
 1 -2 (c_{2}^{2} + c_{3}^{2})   &   -2c_{0}c_{3} + 2c_{1}c_{2}    &   +2c_{0}c_{2} + 2c_{1}c_{3}  \\
 +2c_{0}c_{3} + 2c_{1}c_{2}     &  1 -2 (c_{3}^{2} + c_{1}^{2})   &   -2c_{0}c_{1} + 2c_{2}c_{3}   \\
 -2c_{0}c_{2} + 2c_{1}c_{3}     &   +2c_{0}c_{1} + 2c_{2}c_{3}    &  1 -2 (c_{1}^{2} + c_{2}^{2})
 \end{array} \right | \; .
\label{1.18a}
\end{eqnarray}

Transition  to a general boost is achieved by the change
\begin{eqnarray}
c_{0}  \; \Longrightarrow \;  ch \; {b \over 2}  \;  ,
\; c_{j}  \; \Longrightarrow \;  i\;  sh \; {b \over 2} \;
\; n_{j} \;, \; n_{j}n_{j}=1 \; , \qquad \qquad  O(ib, {\bf n})=
\nonumber
\\
= \left | \begin{array}{rrr}
 1 +  2 sh^{2}(b/2) (n_{2}^{2} + n_{3}^{2})   &    -ish\;b \; n_{3} - 2 sh^{2}(b/2) n_{1}n_{2}    &
   ish\;b \; n_{2} - 2 sh^{2}(b/2) n_{1}n_{3}  \\
  ish\;b \; n_{3} - 2 sh^{2}(b/2) n_{1} n_{2}     &  1 +2 sh^{2}(b/2)(n_{3}^{2} + n_{1}^{2})   &
  -ish\;b \; n_{1} - 2 sh^{2}(b/2) n_{2}n_{3}   \\
 -ish\;b \; n_{2} - 2 sh^{2}(b/2) n_{1}n_{3}     &   ish\;b \; n_{1} -  2 sh^{2}(b/2) n_{2}n_{3}    &
 1 +2 sh^{2}(b/2) (n_{1}^{2} + n_{2}^{2})
 \end{array} \right |
\nonumber
\\
\label{1.20a}
\end{eqnarray}

from this taking in mind  the elementary formula
$
  1- ch\; b = -2 \; sh ^{2}\;(b/2) \; ,
$
we arrive at
\begin{eqnarray}
O(b, {\bf n})
\nonumber
\\
=\left | \begin{array}{rrr}
 1 -  (1-ch\; b) (n_{2}^{2} + n_{3}^{2})        &        -ish\;b \; n_{3} + (1-ch\; b)  n_{1}n_{2}    &
   ish\;b \; n_{2} +  (1-ch\; b)  n_{1}n_{3}  \\
  ish\;b \; n_{3} + (1-ch\; b)  n_{1} n_{2}     &  1 - (1-ch\; b) (n_{3}^{2} + n_{1}^{2})   &
  -ish\;b \; n_{1} +(1-ch\; b)  n_{2}n_{3}   \\
 -ish\;b \; n_{2} + (1-ch\; b)  n_{1}n_{3}     &   ish\;b \; n_{1} +  (1-ch\; b)  n_{2}n_{3}    &
 1  - (1-ch\; b) (n_{1}^{2} + n_{2}^{2})
 \end{array} \right |.
\nonumber
\\
\label{1.20b}
\end{eqnarray}

We need to examine relation
\begin{eqnarray}
\Delta (b, {\bf n})\; ( \; -i \partial_{0} + \alpha^{i}
O_{ij}(b, {\bf n}) \partial_{j} \; ) \; \Psi'
  = \Delta(b, {\bf n}) S J  \; ,
\nonumber
\\
\Delta  = ch\; b - i sh\; b \; n_{1}  \alpha^{1}  -  i sh\; b \;
n_{2}  \alpha^{2} - i sh\; b \; n_{3}  \alpha^{3} \; .
\nonumber
\end{eqnarray}

\noindent
After rather  long   calculation we  can indeed  prove the  general
statement: the matrix Maxwell equation
  is invariant under an arbitrary Lorentzian boost:
\begin{eqnarray}
\Delta (-i\; \partial_{0} + S\alpha^{i}S^{-1} \partial_{i})\; S \Psi =
\Delta SJ \qquad \Longrightarrow \qquad (-i\partial_{0}' +
\alpha^{i} \partial_{i}')\; \Psi' = J'\; ,
\nonumber
\\
S(ib, {\bf n}) = \left | \begin{array}{cc} 1 & 0 \\ 0 & O(ib, {\bf
n})
\end{array} \right |, \qquad \qquad \qquad \qquad
\nonumber
\\
t' = ch\; \beta \; t +   sh\; \beta \; {\bf n}\;   {\bf x} \; ,
\qquad {\bf x}'= +{\bf n}  \; sh\; \beta  \; t + {\bf x}  + (ch\;
\beta -1)\; {\bf n} \; ({\bf n}  {\bf x})\; ,
\nonumber
\\
\partial_{0}' =  ch\; b  \; \partial_{0}  - \;sh\; b\; ( {\bf n}  \nabla )\; , \qquad
\nabla '  =  -sh\; b \; {\bf n}  \; \partial_{0} + [\nabla  +
(ch\;b -1) {\bf n} ({\bf n} \nabla )\; ,
\nonumber
\\
j^{'0} =  ch\; b \; j^{0}  + sh\; b \; ({\bf n} {\bf j}) \;  ,
\qquad {\bf j}' =  +sh\; b \;{\bf n} \;  j^{0} + {\bf j} + (ch\;
b-1)\;  {\bf n}\;  (  {\bf n} {\bf j}) \; .
\label{1.30b}
\end{eqnarray}

\noindent Invariance of the  matrix equation under Euclidean rotations is  achieved in a simpler way:
\begin{eqnarray}
(-i\; \partial_{0} + \alpha^{i} \partial_{i})\; \Psi = J\; , \;
S(a, {\bf n}) = \left | \begin{array}{cc} 1 & 0 \\ 0 & O(a, {\bf
n})
\end{array} \right | \; ,
\; t' = t\; , \qquad {\bf x}'= R(a)  {\bf x}\; ,
\nonumber
\\
(-i\; \partial_{0} + S\alpha^{i}S^{-1} \partial_{i})\; S \Psi = SJ
\qquad \Longrightarrow \qquad (-i\; \partial_{0}' + \alpha^{i}
\partial_{i}')\; \Psi' = J'  \; ,
\nonumber
\\
\partial_{0}' =  \partial_{0} \; , \qquad
\nabla '  =  R(a,-{\bf n}) \nabla \; , \;\;
j^{'0} =   j^{0} \;  , \qquad {\bf j}' =  R(a, {\bf n})  {\bf j}
\; .
\label{1.30d}
\end{eqnarray}

\section{ On  Maxwell equations in a uniform media,  modified Lorentz symmetry }

Let us start with Maxwell equations  in a uniform media:
\begin{eqnarray}
 \mbox{div} \; c{\bf B} = 0 \; , \qquad \mbox{rot}
\;{\bf E} = -{\partial c {\bf B} \over \partial ct} \; ,
\nonumber
\\
\mbox{div}\; {\bf E} = {\rho \over \epsilon
\epsilon_{0}} , \qquad
 \mbox{rot} \; c{\bf B} = \mu \mu_{0}c {\bf J} + \epsilon \mu
   {\partial {\bf E} \over \partial ct} \; .
\label{2.1}
\end{eqnarray}

\noindent The coefficient   $\epsilon \mu\; $  can be factorized as follows
\begin{eqnarray}
\epsilon \mu =  \sqrt{\epsilon \mu } \; \sqrt{\epsilon \mu } = {1
\over k^{2}} \; ,
 \qquad c' = {1 \over  \sqrt{\epsilon_{0} \epsilon \mu_{0} \mu}} = k \; c \;  ;
\nonumber
\end{eqnarray}

\noindent introducing the  variables
\begin{eqnarray}
x^{a} = (x^{0} = kct \; , \;\; x^{i}   ) \; , \qquad  j^{a} =
(j^{0} = \rho , {\bf j} = { {\bf J} \over kc }) ,
\label{2.4}
\end{eqnarray}

\noindent  eqs. (\ref{2.1}) can be rewritten as
\begin{eqnarray}
 \mbox{div} \; k c{\bf B} = 0 \; , \qquad \mbox{rot}
\;{\bf E} = -{\partial k c {\bf B} \over \partial x^{0}} \; ,
\nonumber
\\
\mbox{div}\; {\bf E} = {1\over \epsilon \epsilon_{0}} \; j^{0} ,
\qquad
 \mbox{rot} \; kc{\bf B} = {1 \over \epsilon \epsilon_{0}}\;  {\bf j }  +
   {\partial {\bf E} \over \partial x^{0} } \; ,
\label{2.5a}
\end{eqnarray}

\noindent Equations  (\ref{2.5a}) formally differ from eqs.  (\ref{1.2a}) only in two formal  changes:
$c \Longrightarrow c' = kc$ and $\epsilon_{0} \Longrightarrow \epsilon_{0}\epsilon$ ;
 therefore, all  analysis performed in Section  {\bf 2}
is applicable here:
\begin{eqnarray}
(-i\partial_{0} + \alpha^{i} \partial_{i} )\; \Psi = J \;  , \qquad \psi^{k} =   E^{k} + i  c'B^{k} \; .
\label{2.6c}
\end{eqnarray}

\noindent with the same old  matrices  involved.
The given matrix form of the Maxwell theory in a uniform media
proves existence of symmetry of the theory under  a modified Lorentz group  (Rosen \cite{1952-Rosen}) in which
instead of vacuum speed of light we are to use the modified  speed:
\begin{eqnarray}
c'= k c\; , \qquad c= {1 \over  \sqrt{\epsilon_{0} \mu_{0}}}  \; , \qquad k = {1 \over  \sqrt{\epsilon \mu}} \; .
\label{2.6d}
\end{eqnarray}

\section{On the  matrix form of Maxwell-Minkowski electrodynamics  in  media}

In agreement with Minkowski approach  \cite{1908-Minkowski},
in presence of a uniform media we should  introduce  two electromagnetic tensors
 $F^{ab}$ and  $H^{ab}$ that transform independently under the Lorentz group.
 At this, the known constitutive (or material) relations change their form in the moving reference frame
 (for instance, see \cite{1982-Barykin}).

In the rest media, the Maxwell equations  are
\begin{eqnarray}
\mbox{div} \; {\bf B} = 0 \; , \qquad \mbox{rot} \;{\bf E} =
-{\partial c{\bf B} \over \partial c  t} \; ,
\nonumber
\\
\mbox{div}\; {\bf D} = \rho \; , \qquad
 \mbox{rot} \; { {\bf H} \over  c}  = { {\bf J} \over c }  +
  {\partial {\bf D} \over \partial c t} \; ,
\label{5.1b}
\end{eqnarray}

\noindent with some  constitutive relations.

Quantities with simple transformation laws under the Lorentz group
are
\begin{eqnarray}
{\bf f} = {\bf E} + i c {\bf B}   \; , \qquad  {\bf h} =  {1 \over
\epsilon_{0}} \; ({\bf D} + i {\bf H}  / c )\;  , \;\;
j^{a} = (j^{0}=\rho, \; {\bf j} = {\bf J}/c)\; ;
\label{5.2}
\end{eqnarray}

\noindent where  ${\bf f}, {\bf h}$ are  complex 3-vector under complex orthogonal group $SO(3.C)$, the latter
 is isomorphic  to the  Lorentz  group.
One can combine eqs. (\ref{5.1b})  into following ones
\begin{eqnarray}
 \mbox{div}\; ( {{\bf D}\over \epsilon_{0}}  + i \;c
 {\bf B}) = {1 \over \epsilon_{0}}\; \rho \; ,
\nonumber
\\
- i \partial_{0}  ( {{\bf D}\over \epsilon_{0}}  + i c {\bf B})
+\mbox{rot}\; ( {\bf E} + i{{\bf H}/c \over \epsilon_{0}} ) =
\;{i \over \epsilon_{0}} \;{\bf j} \; .
\label{5.3}
\end{eqnarray}

\noindent
Eqs.  (\ref{5.3})  can be  rewritten in  the form
\begin{eqnarray}
\hspace{40mm} \mbox{div} \; (  { {\bf h } + {\bf h}^{*} \over 2 } + { {\bf f} - {\bf f}^{*} \over 2}) =
{1 \over \epsilon_{0}}\; \rho \; ,
\nonumber
\\
-i\partial_{0} ( { {\bf h } + {\bf h}^{*} \over 2 } + { {\bf f} - {\bf f}^{*} \over 2}) +
\mbox{rot} \; ({ {\bf f } + {\bf f}^{*} \over 2 } + { {\bf h} - {\bf h}^{*} \over 2} )  =
{i \over \epsilon_{0}}\; {\bf j} \; .
\label{5.5}
\end{eqnarray}

\noindent
It has a sense to define two quantities:
\begin{eqnarray}
{\bf M} =   { {\bf h } + {\bf f}  \over 2 } \; , \qquad
{\bf N} =  { {\bf h }^{*} - {\bf f}^{*} \over 2 }  \; ,
\nonumber
\end{eqnarray}

\noindent which  are  different 3-vectors under the group  $SO(3.C)$:
\begin{eqnarray}
{\bf M}' = O \;{\bf M} \; , \qquad  {\bf N}' = O^{*} \;{\bf N} \; .
\nonumber
\end{eqnarray}

\noindent
With respect to Euclidean rotations,  the identity  $O^{*}=O$ holds; whereas for  Lorentzian boosts
we have quite  other identity $O^{*}=O^{-1}$.
In terms of ${\bf M}, {\bf N}$, eqs. (\ref{5.5})  look
\begin{eqnarray}
\mbox{div}\; {\bf M} +  \mbox{div}\; {\bf N} = {1 \over \epsilon_{0}} \;\rho \; ,
\nonumber
\\
-i\partial_{0} {\bf M}  +
\mbox{rot} \; {\bf M}
-i\partial_{0} {\bf N}  -
\mbox{rot} \;{\bf N}
= {i \over \epsilon_{0}} \;{\bf j} \; ,
\label{5.7}
\end{eqnarray}

\noindent or in a matrix form
\begin{eqnarray}
(-i\partial_{0} + \alpha^{i} \partial _{i}) \;  M  + ( -i\partial_{0} + \beta^{i} \partial _{i})  \;  N  = J \; ,
\nonumber
\\
M = \left | \begin{array}{c} 0 \\ {\bf M} \end{array} \right | \;,
 \qquad N = \left | \begin{array}{c} 0 \\ {\bf N} \end{array} \right | \;, \qquad
J = {1 \over \epsilon_{0}} \; \left | \begin{array}{c} \rho  \\ i\; {\bf j} \end{array} \right | \; .
\label{5.8}
\end{eqnarray}

\noindent
The matrices  $\alpha^{i} $  and  $\beta^{i}$  are taken in the form
\begin{eqnarray}
\alpha^{1} = \left | \begin{array}{rrrr}
0 & 1  &  0  & 0  \\
-1 & 0  &  0  & 0  \\
0 & 0  &  0  & -1  \\
0 & 0  &  1  & 0
\end{array}  \right |, \qquad
\alpha^{2} = \left | \begin{array}{rrrr}
0 & 0  &  1  & 0  \\
0 & 0  &  0  & 1  \\
-1 & 0  &  0  & 0  \\
0 & -1  & 0  & 0
\end{array}  \right |, \qquad
\alpha^{3} = \left | \begin{array}{rrrr}
0 & 0  &  0  & 1  \\
0 & 0  & -1  & 0  \\
0 & 1  &  0  & 0  \\
-1 & 0  &  0  & 0
\end{array}  \right |\; ,
\nonumber
\\
\beta^{1} = \left | \begin{array}{rrrr}
0 & 1  &  0  & 0  \\
-1 & 0  &  0  & 0  \\
0 & 0  &  0  & 1  \\
0 & 0  &  -1  & 0
\end{array}  \right |, \qquad
\beta^{2} = \left | \begin{array}{rrrr}
0 & 0  &  1  & 0  \\
0 & 0  &  0  & -1  \\
-1 & 0  &  0  & 0  \\
0 & 1  & 0  & 0
\end{array}  \right |, \qquad
\beta^{3} = \left | \begin{array}{rrrr}
0 & 0  &  0  & 1  \\
0 & 0  & 1  & 0  \\
0 & -1  &  0  & 0  \\
-1 & 0  &  0  & 0
\end{array}  \right | \; .
\nonumber
\end{eqnarray}

\noindent
All of them after squaring give  $-I$, and $\alpha_{i}$ commute with $\beta_{j}$ .

\section{  Minkowski constitutive  relations in a complex 3-vector form
}

Let us  examine how  the constitutive  relations  for an uniform media  behave under  the Lorentz transformations.
One should  start with these  relation in the rest media
\begin{eqnarray}
{\bf D} = \epsilon_{0} \epsilon \; {\bf E} \; ,
\qquad
\; {{\bf H} \over c} = {1 \over \mu_{0} \mu}  {1 \over c^{2}} \; c {\bf B} =
{ \epsilon_{0}\over \mu} \; c{\bf B} \; .
\label{6.1}
\end{eqnarray}

\noindent
Eqs. (\ref{6.1}) can be  rewritten as
\begin{eqnarray}
{{\bf h} + {\bf h}^{*} \over 2} =   \epsilon  \; {{\bf f} + {\bf f}^{*} \over 2} \; ,
\qquad
{{\bf h} - {\bf h}^{*} \over 2} =   {1 \over \mu }  \; {{\bf f} - {\bf f}^{*} \over 2} \; ;
\label{6.2}
\end{eqnarray}

\noindent
from whence it follows
\begin{eqnarray}
2{\bf h} =   (\epsilon +{1 \over \mu }) \; {\bf f}  +
  (\epsilon - {1 \over \mu } ) \; {\bf f}^{*}
\; , \qquad
2{\bf h}^{*} =   (\epsilon +{1 \over \mu }) \; {\bf f}^{*}  +
 (\epsilon - {1 \over \mu } ) \;  {\bf f} \; .
\label{6.3a}
\end{eqnarray}

\noindent
This is a complex form of the constitutive relations  (\ref{6.1}). It should be noted that eqs. (\ref{6.2}) can be
resolved under  ${\bf f} , \; {\bf f}^{*}$ as well:
\begin{eqnarray}
2{\bf f} =  ({1 \over \epsilon} + \mu ) \; {\bf h} +
 ({1 \over \epsilon } - \mu ) \;  {\bf h}^{*} \; ,
\qquad  2{\bf f}^{*} =  ({1 \over \epsilon} + \mu ) \; {\bf h}^{*} +
 ({1 \over \epsilon } - \mu ) \;  {\bf h} \; ;
\label{6.3b}
\end{eqnarray}

\noindent these are the same constitutive equations (\ref{6.3a}) in other  form.
Now let us take into account  the  Lorentz transformations:
\begin{eqnarray}
{\bf f}' =  O \; {\bf f} \; , \qquad  {\bf f}^{'*} =  O ^{*}\; {\bf f}^{*} \; , \qquad  {\bf h}' =  O \;{\bf h} \;,
\qquad {\bf h}^{'*} =  O^{*} \;{\bf h}^{*} \; ;
\nonumber
\end{eqnarray}

\noindent
then eqs. (\ref{6.2}) will become
\begin{eqnarray}
 { O^{-1} {\bf h}' + (O^{-1})^{*}{\bf h}^{*'} \over 2} =
 \epsilon  \; { O^{-1} {\bf f} '+  (O^{-1})^{*} {\bf f}^{*'} \over 2} \; ,
\nonumber
\\
{ O^{-1} {\bf h}' -  (O^{-1})^{*}{\bf h}^{*'} \over 2} =
{1 \over \mu }  \; { O^{-1} {\bf f}' - (O^{-1})^{*}{\bf f}^{*'} \over 2} \; .
\nonumber
\end{eqnarray}

\noindent
Multiplying both equation by  $O$  and summing  (or  subtracting) the results we get
\begin{eqnarray}
2{\bf h}' =  (\epsilon +{1 \over \mu }) \;{\bf f}' +
 (\epsilon - {1 \over \mu } )\; O(O^{-1})^{*} \;{\bf f}^{'*}  \; ,
\nonumber
\\
2{\bf h}^{'*} =   (\epsilon +{1 \over \mu }) \;{\bf f}^{' *}+
(\epsilon - {1 \over \mu } ) \;O^{*}O^{-1} \;{\bf f}^{'} \; .
\label{6.5a}
\end{eqnarray}

\noindent
Analogously, starting from  (\ref{6.3b}) we can produce
\begin{eqnarray}
2{\bf f}' =  ({1 \over \epsilon} + \mu ) \; {\bf h}' +
 ({1 \over \epsilon } - \mu ) \; O(O^{-1})^{*} \; {\bf h}^{'*} \; ,
\nonumber
\\
2{\bf f}^{'*} =  ({1 \over \epsilon} + \mu )  \; {\bf h}^{'*} +
 ({1 \over \epsilon } - \mu ) \;O^{*}O^{-1}  \;{\bf h}' \; .
\label{6.5b}
\end{eqnarray}

Equations   (\ref{6.5a})-(\ref{6.5b}) represent the constitutive relations  after  changing the  reference frame.
In this point one should distinguish between two cases: Euclidean rotation and Lorentzian boosts.
Indeed, for any  Euclidean rotations
\begin{eqnarray}
O^{*} = O \; , \qquad \Longrightarrow \qquad O(O^{-1})^{*}  = I \; , \qquad  O^{*}O^{-1} = I \; ;
\nonumber
\end{eqnarray}

\noindent
and therefore   eqs.  (\ref{6.5a})-(\ref{6.5b}) take the  form of  (\ref{6.3a})-(\ref{6.3b}); in other words, at Euclidean rotations
the constitutive  relations do not change their form.
However, for any pseudo-Euclidean rotations (Lorentzian boosts)
\begin{eqnarray}
O^{*} = O^{-1} , \qquad \Longrightarrow \qquad O(O^{-1})^{*}  = 0^{2}, \qquad  O^{*}O^{-1} = O^{*2} \; ;
\nonumber
\end{eqnarray}

\noindent  and  eqs. (\ref{6.5a})-(\ref{6.5b})  look
\begin{eqnarray}
2{\bf h}' =   (\epsilon +{1 \over \mu })\; {\bf f}' +
 (\epsilon - {1 \over \mu } ) \;O^{2} \; {\bf f}^{'*}  \; ,
\nonumber
\\
2{\bf h}^{'*} =  (\epsilon +{1 \over \mu })\; {\bf f}^{' *}+
 (\epsilon - {1 \over \mu } ) \;O^{2} \;{\bf f}^{'} \; ;
\label{6.6a}
\\
2{\bf f}' =  ({1 \over \epsilon} + \mu ) \; {\bf h}' +
 ({1 \over \epsilon } - \mu ) \; O^{*2}  \;{\bf h}^{'*} \; ,
\nonumber
\\
2{\bf f}^{'*} = ({1 \over \epsilon} + \mu )  \; {\bf h}^{'*} +
({1 \over \epsilon } - \mu ) \;O^{*2} \; {\bf h}' \; .
\label{6.6b}
\end{eqnarray}

In complex 3-vector  form these relations seem to be shorter than in real 3-vector form:
\begin{eqnarray}
2{\bf D}'  =  \epsilon_{0} \epsilon  \;  [ \;  ( I  + {O O  + O^{*} O^{*} \over 2} )\; {\bf  E}'\;
+ {O O  - O^{*} O^{*} \over 2i} \;   c{\bf  B}'\; ] \;  +
\nonumber
\\
+  {\epsilon_{0} \over \mu }  \; [ \;( I  - {O O  + O^{*} O^{*} \over 2} ) \;   {\bf  E}'\;
- {OO  - O^{*} O^{*} \over 2i } \;  c {\bf  B}'\; ] \; ,
\nonumber
\\
2{\bf H}' /c =  \epsilon_{0} \epsilon  \;  [ \; (I  - {OO + O^{*} O^{*} \over 2} ) \;c {\bf  B}'\;
+ {OO  - O^{*} O^{*} \over 2i} \;   {\bf  E}'\; ] \;  +
\nonumber
\\
+ {\epsilon_{0} \over \mu }  \; [ \; (I + {OO  + O^{*} O^{*} \over 2} )\;   c{\bf B}'\;
- {O O  - O^{*} O^{*} \over 2i } \;   {\bf  E}'\; ] \; .
\label{6.8a}
\end{eqnarray}

\noindent They can be written differently
\begin{eqnarray}
{\bf D}'  =  {\epsilon_{0}  \over 2 } \;  \left \{
[\; ( \epsilon  \; + {1 \over \mu })  +  ( \epsilon  \; -{1 \over \mu }) \; \mbox{Re} \; O^{2} \;  ]\;  {\bf  E}'\;
+ ( \epsilon  \; - {1 \over \mu })   \; \mbox{Im} \; O^{2} \;  c{\bf  B}'\; ] \;  \right  \}\; ,
\nonumber
\\
 {{\bf H}' \over c } =  {\epsilon_{0} \over 2}   \; \left \{
 [
 ( \epsilon  \; + {1 \over \mu })  -  ( \epsilon  \; -{1 \over \mu }) \; \mbox{Re} \; O^{2} \;  ]\;  c{\bf  B}'\;
+ ( \epsilon  \; - {1 \over \mu })   \; \mbox{Im} \; O^{2} \;  {\bf  E}'\; ] \;  \right  \} \; .
\label{6.8b}
\end{eqnarray}

The matrix $O^{2}$  can be presented differently with the help of double angle variable:
\begin{eqnarray}
O^{2}=
\nonumber
\\
\left | \begin{array}{ccc}
ch\;2b  + (1 -ch\; 2b )\;  n_{1}^{2}   &
(1 -ch\; 2b ) \;  n_{1}n_{2}   -  i  sh\;2 b  \;  n_{3}   &
(1 -ch\; 2b ) \;  n_{3}n_{1}   +  i   \; sh\;2b \; n_{2}  \\[2mm]
(1 -ch\; 2b )\;  n_{1}n_{2}   +  i   \; sh\;2b \; n_{3} &
ch\;2b  + (1 -ch\; 2b )\;  n_{2}^{2}  &
(1 -ch\; 2b ) \;  n_{2}n_{3}  -  i  \; sh\;2b \; n_{1} \\[2mm]
(1 -ch\; 2b ) \;  n_{3}n_{1}   -  i   \; sh\;2b \; n_{2}        &
(1 -ch\; 2b ) \;  n_{2}n_{3}   +  i   \; sh\;2b \; n_{1} &
ch\;2b  + (1 -ch\; 2b )\;  n_{3}^{2}
\end{array} \right |.
\nonumber
\\
\label{6.14}
\end{eqnarray}

The previous result can be easily extended to more generale medias, let us restrict  ourselves to linear medias.
Indeed, arbitrary linear media is characterized by the following constitutive equations:
\begin{eqnarray}
{\bf D}= \epsilon_{0} \;\epsilon(x) \; {\bf E}+ \epsilon_{0}c
\;\alpha(x) \; {\bf B} \; ,
\qquad
{\bf H}= \epsilon_{0}c \;\beta(x) \; {\bf E}+ {1 \over \mu_{0}}
\;\mu(x) \; {\bf B} \; ,
\label{6.16}
\end{eqnarray}

\noindent where $\epsilon(x), \mu(x), \alpha(x), \beta(x)$ are $3 \times 3$  dimensionless matrices.
Eqs. (\ref{6.16})  should be rewritten in terms of complex vectors ${\bf f}, {\bf h}$:
\begin{eqnarray}
{{\bf h} + {\bf h}^{*} \over 2}  = \epsilon(x) \; {{\bf f} + {\bf f}^{*} \over 2} +
\alpha(x) \; {{\bf f} - {\bf f}^{*} \over 2i }\; ,
\nonumber
\\
{{\bf h} - {\bf h}^{*} \over 2i}  = \beta(x) \;  {{\bf f} + {\bf f}^{*} \over 2} +
\mu(x) \; {{\bf f} - {\bf f}^{*} \over 2i  } \; .
\label{6.16b}
\end{eqnarray}

\noindent
From (\ref{6.16}) it follows
\begin{eqnarray}
{\bf h}=
  [ \; (\epsilon (x) + \mu  (x))  +i (\beta (x) - \alpha(x)) \;  ]  \; {\bf f} +
 [\;
(\epsilon(x) - \mu(x) )  +i (\beta(x) + \alpha (x) ) \;  ] \;  {\bf f}^{*}  \; ,
\nonumber
\\
{\bf h}^{*}=
 [ \;
(\epsilon (x) + \mu  (x))  -i (\beta (x)  - \alpha (x)) \;  ] \;  {\bf f}^{*} +
[ \;
(\epsilon (x)  - \mu  (x) )  -i (\beta  (x) + \alpha (x) ) \;  ]  \; {\bf f} \;  .
\label{6.17}
\end{eqnarray}

Under Lorentz transformations,  relations (\ref{6.17}) will take the form
\begin{eqnarray}
O^{-1} {\bf h} ' =
 \left [
(\epsilon  (x) + \mu  (x))  +i (\beta(x) - \alpha (x)) \right ] O^{-1} {\bf f}'\;  +
\nonumber
\\
\left [
(\epsilon (x) - \mu (x) )  +i (\beta (x) + \alpha (x)) \right ]  (O^{-1})^{*}{\bf f}^{'*}  ,
\nonumber
\\
(O^{-1})^{*}{\bf h}^{'*}=
 \left [
(\epsilon (x) + \mu (x))  -i (\beta (x) - \alpha (x)) \right ]  (O^{-1})^{*} {\bf f}^{'*} +
\nonumber
\\
\left [
(\epsilon (x) - \mu (x) )  -i (\beta (x) + \alpha (x)) \right ]  (O^{-1}) {\bf f}'  ,
\nonumber
\end{eqnarray}

\noindent or
\begin{eqnarray}
 {\bf h} ' =
\epsilon_{0} \left [
(\epsilon  (x)+ \mu (x) )  +i (\beta  (x) - \alpha (x) ) \right ]  {\bf f}' +
 \left [ \;
(\epsilon (x) - \mu  (x))  +i (\beta (x) + \alpha (x) ) \right ]  [O(O^{-1})^{*}] \; {\bf f}^{'*} ,
\nonumber
\\
{\bf h}^{'*}=
\epsilon_{0} \left [ \;
(\epsilon  (x) + \mu (x) )  -i (\beta (x) - \alpha (x)) \right ] \;   {\bf f}^{'*} +
\left [
(\epsilon (x) - \mu (x) )  -i (\beta  (x) + \alpha (x)) \right ]  [O^{*}(O^{-1}) ] \; {\bf f}' .
\nonumber
\end{eqnarray}

\noindent For Euclidean rotation, the constitutive relations
preserve  their form. For Lorentz boosts we have \begin{eqnarray} {\bf h} ' =
 \left [ \;
(\epsilon  (x)+ \mu (x) )  +i\; (\beta  (x) - \alpha (x)) \; \right ]  {\bf f}' +
\left [\;
(\epsilon (x) - \mu (x) )  +i \;(\beta  (x) + \alpha (x) ) \; \right ]  O^{2}  {\bf f}^{'*} \; ,
\nonumber
\\
{\bf h}^{'*}=
 \left [ \;
(\epsilon (x) + \mu (x) )  -i \;(\beta (x) - \alpha (x)) \; \right ]   {\bf f}^{'*} +
\left [\;
(\epsilon (x) - \mu (x))  -i \; (\beta  (x)+ \alpha (x))\;  \right ]  O^{*2} {\bf f}' .
\nonumber
\\
\label{6.20}
\end{eqnarray}

\noindent They  are the constitutive  equations for arbitrary linear medias in a moving reference frame
(similar  formulas were produced in quaternion  formalism in  \cite{1985-Berezin}-\cite{1992-Berezin}).

\section{Symmetry properties of the matrix  Maxwell equation  in a uniform  media }

As noted, Maxwell equations  in any media can be presented  in the matrix form  as follows:
\begin{eqnarray}
(-i\partial_{0} + \alpha^{i} \partial _{i}) \;M  + ( -i\partial_{0}
+ \beta^{i} \partial _{i}) \; N  = J \; .
\label{7.3b}
\end{eqnarray}

\noindent We  are to study symmetry properties of this equation under complex rotation group SO(3.C).
 The  terms with  $\alpha^{j}$  matrices  were  examined in Section {\bf 1}),
 the terms with  $\beta^{j}$ matrix is  new.
We restrict ourselves to demonstrating  the Lorentz symmetry of eq. (\ref{7.3b}) under two simplest
transformations.

First, let  us consider  the Euclidean rotation in the plane $(1-2)$;  examine additionally only
the term with $\beta$-matrices:
\begin{eqnarray}
S\beta^{1}S^{-1} =
 \cos a \; \beta^{1} - \sin a  \; \beta^{2} = \beta^{j} O_{j1} \; ,
\nonumber
\\
\beta^{2}S^{-1} =
 \sin a\; \beta^{1} + \cos a \; \beta^{2} = \beta^{j} O_{j2}\; ,
\nonumber
\\
S \beta^{3}S^{-1} =  \beta^{3}   = \beta^{j} O_{j3} \; .
\label{7.5c}
\end{eqnarray}

\noindent
Therefore, we conclude that eq.  (\ref{7.3b})  is symmetrical under Euclidean rotations
in accordance with the relations
\begin{eqnarray}
(-i\partial_{0} + S \alpha^{i}S^{-1} \partial _{i})\; M'  +
(-i\partial_{0} + S \beta^{i} S^{-1} \partial _{i}) \; N' = + S J
\; , \;\;\; \Longrightarrow
\nonumber
\\
(-i\partial_{0} +  \alpha^{i}  \partial' _{i}) \; M' +
(-i\partial_{0} +  \beta^{i}  \partial ' _{i}) \; N'   = +  J' \; .
\;
\label{7.6}
\end{eqnarray}

For the Lorentz boost in the plane $(0-3)$ we have
\begin{eqnarray}
M' = S M \; , \qquad N' = S^{*} N = S^{-1} N , \qquad  S^{*} =
S^{-1} \; ;
\nonumber
\end{eqnarray}

\noindent and eq.  (\ref{7.3b}) takes the form (note that the additional transformation
 $\Delta = \Delta_{(\alpha)}$  is combined  in terms of  $\alpha^{j}$ (see Sec. {\bf 2})
\begin{eqnarray}
\Delta_{(\alpha )}  S \; \left [ \; (-i\partial_{0} + \alpha^{i}
\partial _{i})\; S^{-1} M'   + ( -i\partial_{0} + \beta^{i}
\partial _{i})\;S  N' \;  \right ]  =  \Delta S J \; ,
\nonumber
\end{eqnarray}

\noindent or
\begin{eqnarray}
\Delta _{(\alpha )} \; \left [ (-i\partial_{0} + S\alpha^{i}
S^{-1}  \partial _{i})\;  M'   + S^{2} ( -i\partial_{0}   +
S^{-1} \beta^{i} S\partial _{i})\; N' \right ]  = J' \; ,
\nonumber
\end{eqnarray}

\noindent and further
\begin{eqnarray}
(-i\partial_{0}' + \alpha^{i}  \partial' _{i})\;  M'   +
\Delta_{(\alpha )} S^{2} ( -i\partial_{0}   +  S^{-1} \beta^{i}
S\partial _{i})\; N'   = J' \; .
\label{7.7c}
\end{eqnarray}

\noindent It remains to prove the relationship
\begin{eqnarray}
\Delta _{(\alpha )} S^{2} \; ( -i\partial_{0}   +  S^{-1}
\beta^{i} S\partial _{i})\; N' = ( -i\partial_{0}'   +   \beta^{i}
\partial _{i} ')\; N' \; .
\label{7.8}
\end{eqnarray}

\noindent  By simplicity reason one may expect two  identities:
\begin{eqnarray}
\Delta _{(\alpha )} S^{2}= \Delta _{(\beta )}  \qquad
\Longleftrightarrow \qquad \Delta _{(\alpha )} S = \Delta _{(\beta
)} S^{-1} \; ,
 \label{7.9a}
\end{eqnarray}
and
\begin{eqnarray}
\Delta _{(\beta )} ( -i\partial_{0}   +  S^{-1} \beta^{i}
S\partial _{i})\; N' = ( -i\partial_{0} '  +   \beta^{i} \partial
_{i}')\; N'  \; .
\label{7.9b}
\end{eqnarray}

Let us prove them for a Lorentzian boost in the plane $0-3$:
\begin{eqnarray}
S= \left | \begin{array}{cccc}
1  &       0   &      0  &  0  \\
0  &  ch \; b   &  -i \; sh\; b  &  0  \\
0  & i\; sh \; b   & ch \; b  &  0  \\
0 & 0  & 0  &  1
\end{array} \right | \; , \qquad
S^{-1} = \left | \begin{array}{cccc}
1  &       0   &      0  &  0  \\
0  &  ch \; b   &  -i \; sh\; b  &  0  \\
0  & i\; sh \; b   & ch \; b  &  0  \\
0 & 0  & 0  &  1
\end{array} \right | \; ;
\nonumber
\end{eqnarray}

\noindent we readily  get
\begin{eqnarray}
S^{-1} \beta^{1}S =  ch \; b \; \beta^{1} - i \; sh \; b  \;
\beta^{2} = \beta^{j} O^{-1}_{j1} \; ,
\nonumber
\\
S^{-1} \beta^{2}S =
  i \; sh\; b  \beta^{1} + ch\; b  \; \beta^{2} = \beta^{j} O^{-1}_{j2}\; ,
\;\; S^{-1} \beta^{3 }S = \beta^{3} =  \beta^{j} O^{-1}_{j3}\; .
\label{7.10}
\end{eqnarray}

\noindent To verify identity
$
\Delta _{(\alpha )} S = \Delta _{(\beta )} S^{-1} \; ,
$
or
\begin{eqnarray}
(ch\; b - i sh\; b \; \alpha^{3}) S = (ch\; b - i sh\; b \;
\beta^{3}) S^{-1} \; ,
\nonumber
\end{eqnarray}

\noindent let us calculate  separately the left and right  parts:
\begin{eqnarray}
(ch\; b - i sh\; b \; \alpha^{3}) S =
 (ch\; b - i sh\; b \; \beta^{3}) S^{-1} =
 \left | \begin{array}{cccc}
ch\; b  &   0   &   0  &  -i\; sh\; b  \\
0  &   1   &   0  &  0  \\
0  &   0   &   1  &  0  \\
i\; sh\; b   & 0  & 0  &  ch\; b
\end{array} \right | \; .
\nonumber
\end{eqnarray}

\noindent they coincide with each other, so eq.  (\ref{7.9a})  holds.
It remains to prove relation   (\ref{7.9b}). Allowing for the properties of $\beta$--matrices
\begin{eqnarray}
(\beta^{0})^{2} = -I , \qquad (\beta^{1}) ^{2} = -I , ... \qquad
\beta ^{1} \beta ^{2} =  - \; \beta^{3} ,  \qquad \beta ^{2} \beta
^{1} =  + \; \beta^{3} ...
\nonumber
\end{eqnarray}

\noindent we readily  find
\begin{eqnarray}
\Delta _{(\beta )} \; ( -i\partial_{0}   +  S^{-1} \beta^{i}
S\partial _{i})\; N' =
 (ch\; b - i sh\; b \; \beta^{3}) \; [\; -i \partial_{0} + \beta^{3} \partial_{3} +
\nonumber
\\
+   (ch \; b \; \beta^{1} - i \; sh \; b  \; \beta^{2})  \;
\partial_{1} +
   ( i \; sh\; b  \beta^{1}  +  ch\; b  \; \beta^{2} ) \; \partial_{2} \; ] \; N' =
\nonumber
\\
= [\; -i (ch\;b \; \partial_{0} - sh\; b \; \partial_{3}) +
\beta^{3} \; (-sh\; b \; \partial_{0} +
 ch\; b \; \partial_{3}) + \beta^{1} \; \partial_{1} + \beta^{2}\;  \partial_{2} \;] \; N' \; ,
\nonumber
\end{eqnarray}

\noindent that is
\begin{eqnarray}
\Delta _{(\beta )} ( -i\partial_{0}   +  S^{-1} \beta^{i}
S\partial _{i})\; N' = ( -i\partial_{0}'    +   \beta^{1} \partial
_{1} + \beta^{2} \partial _{2} + \beta^{3} \partial' _{3} )\; N'
\; ;
\label{7.11b}
\end{eqnarray}

\noindent the  relation  (\ref{7.9b}) holds.
Thus, the symmetry of the matrix Maxwell equation in media  under the Lorentz group  is proved.

\section{Maxwell theory, Dirac matrices and electromagnetic 4-vectors}

Let us  shortly discuss two points relevant to the above  matrix formulation of the Maxwell theory.

First, let write down  explicit form for Dirac matrices in spinor basis:
\begin{eqnarray}
\gamma^{0} = \left | \begin{array}{cccc}
0 & 0 & 1 & 0 \\
0 & 0 & 0 & 1 \\
1 & 0 & 0 & 0 \\
0 & 1 & 0 & 0
\end{array} \right | , \qquad \gamma^{5} = -i \gamma^{0} \gamma^{1} \gamma^{2} \gamma^{3} =
\left | \begin{array}{cccc}
-1 & 0 & 0 & 0 \\
0 & -1 & 0 & 0 \\
0 & 0 & 1 & 0 \\
1 & 0 & 0 & 1
\end{array} \right | ,
\nonumber
\\
\gamma^{1} = \left | \begin{array}{cccc}
0 & 0 & 0 & -1 \\
0 & 0 & -1 & 0 \\
0 & 1 & 0 & 0 \\
1 & 0 & 0 & 0
\end{array} \right | ,
\gamma^{2} = \left | \begin{array}{cccc}
0 & 0 & 0 & i \\
0 & 0 & -i & 0 \\
0 & -i & 0 & 0 \\
i & 0 & 0 & 0
\end{array} \right | ,
\gamma^{3} = \left | \begin{array}{cccc}
0 & 0 & -1 & 0 \\
0 & 0 & 0 & 1 \\
1 & 0 & 0 & 0 \\
0 & -1 & 0 & 0
\end{array} \right | .
\nonumber
\end{eqnarray}

Taking in mind expressions  for $\alpha^{i}, \beta^{i}$, we immediately  see the identities
\begin{eqnarray}
\alpha^{1}  =i \gamma^{0} \gamma^{2} , \qquad
\alpha^{2}  =  \gamma^{0} \gamma^{5} , \qquad
\alpha^{3}  = i \gamma^{5} \gamma^{2} ,
\nonumber
\\
\beta^{1} = - \gamma^{3} \gamma^{1} , \qquad \beta^{2} = - \gamma^{3} , \qquad \beta^{3} = - \gamma^{1}
\label{a2}
\end{eqnarray}

\noindent
so the Maxwell matrix equation in media takes  the form
\begin{eqnarray}
(-i \partial_{0}  + i \gamma^{0} \gamma^{2} \partial_{1} + \gamma^{0} \gamma^{5} \partial_{2}+
i \gamma^{5} \gamma^{2} \partial_{3} ) \; M \; +
\nonumber
\\
+ \; ( -i \partial_{0} - \gamma^{3} \gamma^{1} \partial_{1} - \gamma^{3} \partial_{2} - \gamma^{1} \partial_{3} ) \; N = J
\label{a3}
\end{eqnarray}

\noindent
This Dirac matrix-based form does not seem to be  very useful to apply in the  Maxwell theory,  it
does not prove much similarity with ordinary Dirac equation
(though  that  analogy   was often discussed in the literature).

Now starting from
 electromagnetic 2-tensor and dual to it:
\begin{eqnarray}
\tilde{F}_{\rho\sigma}= {1 \over 2} \; \epsilon_{\rho\sigma\alpha \beta} F^{\alpha \beta} \; , \qquad
F_{\alpha\beta}=  -{1 \over 2} \; \epsilon_{\alpha \beta \rho \sigma} \tilde{F} ^{\rho \sigma}
\nonumber
\end{eqnarray}

\noindent let us introduce two  electromagnetic 4-vectors (below $u^{\alpha}$ is any 4-vector that in general must
not coincide with 4-velocity)
\begin{eqnarray}
e^{\alpha} = u_{\beta} F^{\alpha \beta} \; , \qquad b^{\alpha} = u_{\beta} \tilde{F}^{\alpha \beta} \;  ,
\qquad u^{\alpha} u_{\alpha} = 1 \; ;
\label{a4}
\end{eqnarray}

\noindent
inverse formulas are
\begin{eqnarray}
F^{\alpha \beta}  =  (e^{\alpha} \; u^{\beta} - e^{\beta} \;  u^{\alpha}) -
\epsilon^{\alpha \beta \rho \sigma} \; b_{\rho} \; u_{\sigma} \; ,
\nonumber
\\
\tilde{F}^{\alpha \beta}  = ( b^{\alpha} \; u^{\beta} - b^{\beta} \; u^{\alpha} ) +
\epsilon^{\alpha \beta \rho \sigma} \; e_{\rho} \;  u_{\sigma} \; .
\label{a5}
\end{eqnarray}

\noindent Such electromagnetic 4-vector are presented always  in the literature   on the electrodynamics of moving bodies,
from the very beginning of
relativistic tensor form of electrodynamics -- see Minkowski \cite{1908-Minkowski}, Gordon \cite{1923-Gordon},
Mandel'stam -- Tamm \cite{1925-Mandel'stam}-\cite{1925-Tamm(1)}-\cite{1925-Tamm(2)}; for instance see
Y\'epez --  Brito -- Vargas \cite{1988-Nunez}. The interest to these  field variables  gets renewed
after Esposito paper \cite{1998-Esposito} in 1998.

In 3-dimensional notation
\begin{eqnarray}
E^{1} = -E_{1} = F^{10} \;  ,  \qquad cB^{1} =cB_{1} = \tilde{F}^{10} = -F_{23} \;  , \qquad \mbox{and so on}
\nonumber
\end{eqnarray}

\noindent
the formulas  (\ref{a4})  take the form
\begin{eqnarray}
e^{0} = {\bf u} \; {\bf E} \;  , \qquad {\bf e} = u^{0}\;  {\bf E} + c\; {\bf u} \times {\bf B} \; ,
\nonumber
\\
b^{0} = c\; {\bf u} \; {\bf B} \;  , \qquad {\bf b} = c\; u^{0}\;  {\bf B} - {\bf u} \times {\bf E} \; ,
\label{a6}
\end{eqnarray}

\noindent
or in symbolical form
\begin{eqnarray}
(e, b) = U(u) \; ({\bf E}, {\bf B}) \; ;
\nonumber
\end{eqnarray}

\noindent
and inverse the formulas  (\ref{a5})  look
\begin{eqnarray}
{\bf E} = {\bf e} \; u^{0} - e^{0} \;  {\bf u}   + {\bf b} \times {\bf u}
\; ,
\nonumber
\\
c\; {\bf B} = {\bf b} \; u^{0} - b^{0} \;  {\bf u}   - {\bf e} \times {\bf u} \; .
\label{a7}
\end{eqnarray}

\noindent
or in symbolical form
\begin{eqnarray}
({\bf E}, {\bf B}) = U^{-1}(u) \;  (e, b)  \; ;
\nonumber
\end{eqnarray}

\noindent
The relationships can be  checked  by  direct calculation:
\begin{eqnarray}
{\bf E} = ( u^{0}\;  {\bf E} + c\; {\bf u} \times {\bf B}) \; u^{0} - ({\bf u} \; {\bf E}) \;  {\bf u}   +
(c\; u^{0}\;  {\bf B} - {\bf u} \times {\bf E}) \times {\bf u} =
\nonumber
\\
= u^{0}  u^{0}\;  {\bf E} +  c\; u^{0} ({\bf u} \times {\bf B}) - ({\bf u} \; {\bf E}) \;  {\bf u}  +
c\; u^{0} ({\bf B} \times {\bf u}) + {\bf u } \times ( {\bf u} \times {\bf E})
= (u^{0}u^{0} -{\bf u}{\bf u}) {\bf E} ={\bf E} \; ,
\nonumber
\\
c{\bf B} = ( c\; u^{0}\;  {\bf B} - {\bf u} \times {\bf E}) \; u^{0} -
(c\;{\bf u} \; {\bf B})  \;  {\bf u}   - (u^{0}\;  {\bf E} + c\; {\bf u} \times {\bf B})      \times {\bf u}=
\nonumber
\\
=  c\; u^{0}u^{0} \;  {\bf B} - u^{0} ({\bf u} \times {\bf E}) - c\; ({\bf u} \; {\bf B})  \;  {\bf u}
- u^{0} ({\bf E} \times {\bf u}) + c\; {\bf u } \times ( {\bf u} \times {\bf B})
 =
 c \; (u^{0}u^{0} -{\bf u}{\bf u}) {\bf B} = c \; {\bf B} \; .
\nonumber
\end{eqnarray}

The above  possibility is often used to produce a special form of the Maxwell equations. For simplicity,  let
us consider the
vacuum case:
\begin{eqnarray}
\partial_{\alpha} F_{\beta \gamma} + \partial_{\beta} F_{\gamma \alpha} +
\partial_{\gamma} F_{\alpha \beta} = 0\; , \qquad
\partial_{\alpha}F^{\alpha \beta} = \epsilon_{0}^{-1} j^{\beta}
\nonumber
\end{eqnarray}

\noindent or differently with the help of the dual tensor:
\begin{eqnarray}
\partial_{\beta}\tilde{F}^{\beta \alpha} = 0
\; , \qquad
\partial_{\alpha}F^{\alpha \beta } = \epsilon_{0}^{-1} j^{\beta}\;
\label{a8}
\end{eqnarray}

\noindent
They  can  be transformed to variables $b^{\alpha}, b^{\alpha}$:
\begin{eqnarray}
\partial_{\alpha} ( b^{\alpha} \; u^{\beta} - b^{\beta} \; u^{\alpha}  +
\epsilon^{\alpha \beta \rho \sigma} \; e_{\rho} \;  u_{\sigma} ) = 0
\; ,
\nonumber
\\
\partial_{\alpha} ( e^{\alpha} \; u^{\beta} - e^{\beta} \;  u^{\alpha} -
\epsilon^{\alpha \beta \rho \sigma} \; b_{\rho} \; u_{\sigma} ) = \epsilon_{0}^{-1} j^{\beta}\; .
\nonumber
\end{eqnarray}

They can be combined into equations for complex field function
\begin{eqnarray}
\Phi^{\alpha} = e^{\alpha}+ ib^{\alpha}\; ,
\qquad
\partial_{\alpha} \; [ \; \Phi^{\alpha}  u^{\beta} -   \Phi^{\beta} u^{\alpha} +
i \epsilon^{\alpha \beta \rho \sigma} \Phi _{\rho} u_{\sigma}\; ]=
\epsilon_{0}^{-1} j^{\beta}
\nonumber
\end{eqnarray}

\noindent or differently
\begin{eqnarray}
\partial_{\alpha} \; [ \;  \delta^{\alpha}_{\gamma} u^{\beta}   -
  \delta^{\beta}_{\gamma} u^{\alpha}   +
i \epsilon^{\alpha \beta \rho \sigma}  g_{\rho \gamma}  u_{\sigma}\; ]\; \Phi^{\gamma} =
\epsilon_{0}^{-1} j^{\beta}
\label{a10}
\end{eqnarray}

\noindent
This is Esposito's representation \cite{1998-Esposito} of the Maxwell equations.
One may introduce four matrices, functions of  4-vector $u$:
\begin{eqnarray}
(\Gamma^{\alpha})^{\beta}_{\;\;\gamma}  =   \delta^{\alpha}_{\gamma}  u^{\beta}  -
  \delta^{\beta}_{\gamma} u^{\alpha}   +
i \epsilon^{\alpha \beta \rho \sigma}  g_{\rho \gamma}  u_{\sigma}\; ,
\label{a11}
\end{eqnarray}

\noindent
then eq. (\ref{a10}) becomes
\begin{eqnarray}
\partial_{\alpha} (\Gamma^{\alpha})^{\beta}_{\;\;\gamma} \; \Phi^{\gamma} =
\epsilon_{0}^{-1} j^{\beta} \; , \qquad  \mbox{or}\qquad \Gamma^{\alpha}\partial_{\alpha}  \Phi =  \epsilon_{0}^{-1} j  \; .
\label{a12}
\end{eqnarray}

In the 'rest reference frame'  when $u^{\alpha}= (1, 0,0,0)$,
the matrices $\Gamma^{\alpha}$ become simpler and $\Phi =\Psi$:
\begin{eqnarray}
\Gamma^{0}  =    \left | \begin{array}{cccc}
  0 & 0 & 0 & 0  \\
  0 & -1 &0 & 0  \\
  0 & 0 & -1& 0  \\
  0 & 0 & 0& -1
\end{array} \right |
     ,
\Gamma^{1}=
 \left | \begin{array}{cccc}
  0 & 1 & 0 & 0  \\
  0 & 0 & 0 & 0  \\
  0 & 0 & 0 & i  \\
  0 & 0 & -i& 0
\end{array} \right | ,
\Gamma^{2}=
 \left | \begin{array}{cccc}
  0 & 0 & 1 & 0  \\
  0 & 0 & 0 & -i  \\
  0 & 0 & 0 & 0  \\
  0 & i & 0& 0
\end{array} \right |,
\Gamma^{3}=
 \left | \begin{array}{cccc}
  0 & 0 & 0 & 1  \\
  0 & 0 & i &   \\
  0 & -i & 0 & 0  \\
  0 & 0 & 0& 0
\end{array} \right | \; .
\label{a13}
\end{eqnarray}

\noindent
and eq. (\ref{a12}) takes the form
\begin{eqnarray}
\left | \begin{array}{rrrr}
0  &  \partial_{1}  &    \partial_{2}  &  \partial_{3} \\
0  & -\partial_{0}  &  i\partial_{3}  & -i\partial_{2}  \\
0  & -i\partial_{3}  &   -\partial_{0}  &  i\partial_{1}  \\
0  & i\partial_{2}  &  -i\partial_{1}  & -\partial_{0}
\end{array} \right |
\left | \begin{array}{c}
0 \\ E^{1} + icB^{1} \\ E^{2} + icB^{2} \\ E^{3} + icB^{3}
\end{array} \right | =
\epsilon^{-1}_{0}
\left | \begin{array}{c}
\rho  \\ j^{1}  \\ j^{2}  \\ j^{3}
\end{array} \right | \; ,
\label{a14}
\end{eqnarray}

\noindent
or
\begin{eqnarray}
\mbox{div} \; ({\bf E} + ic{\bf B})  = \epsilon^{-1}_{0} \; \rho\; ,\qquad
-\partial_{0}  ({\bf E} + ic{\bf B}) - i \; \mbox{rot} \; ({\bf E} + ic{\bf B}) = \epsilon^{-1}_{0}
  {\bf j} \; .
\label{a15}
\end{eqnarray}

\noindent
From whence we get equations
\begin{eqnarray}
 \mbox{div} \; c{\bf B} = 0 \; , \qquad \mbox{rot}
\;{\bf E} = -{\partial c {\bf B} \over \partial ct} \; ,
\nonumber
\\
\mbox{div}\; {\bf E} = {\rho \over  \epsilon_{0}} , \qquad
 \mbox{rot} \; c{\bf B} =  {{\bf j} \over \epsilon_{0}}  +
   {\partial {\bf E} \over \partial ct} \; ,
\nonumber
\end{eqnarray}

\noindent which coincides with eqs. (\ref{1.2a}).
Relation (\ref{a13}) corresponds to a special choice in (\ref{1.4}):
\begin{eqnarray}
(-i\alpha^{0} \partial_{0} + \alpha^{j} \partial_{j} ) \Psi = {1 \over  \epsilon_{0}}
\left | \begin{array}{c}
 \rho  \\  ij^{1} \\ i j^{2} \\ ij^{3}
\end{array} \right |
 \;
, \qquad \Psi = \left | \begin{array}{c} 0 \\ E^{1}+ icB^{1} \\ E^{2} +icB^{2}
\\ E^{3} +i cB^{3}
\end{array} \right | \; ,
\nonumber
\\
\alpha^{0} = \left | \begin{array}{rrrr}
0 & 0  &  0  & 0  \\
0 & 1  &  0  & 0  \\
0 & 0  &  1  & 0  \\
0 & 0  &  0  & 1
\end{array}  \right | \; ,
\alpha^{1} = \left | \begin{array}{rrrr}
0 & 1  &  0  & 0  \\
0 & 0  &  0  & 0  \\
0 & 0  &  0  & -1  \\
0 & 0  &  1  & 0
\end{array}  \right |\;  ,
\alpha^{2} = \left | \begin{array}{rrrr}
0 & 0  &  1  & 0  \\
0 & 0  &  0  & 1  \\
0 & 0  &  0  & 0  \\
0 & -1  & 0  & 0
\end{array}  \right |\;  ,
\alpha^{3} = \left | \begin{array}{rrrr}
0 & 0  &  0  & 1  \\
0 & 0  & -1  & 0  \\
0 & 1  &  0  & 0  \\
0 & 0  &  0  & 0
\end{array}  \right | \; ;
\nonumber
\\
\label{a16}
\end{eqnarray}

\noindent
and relations (\ref{a13}) -- (\ref{a14}) are referred to  (\ref{a16})  trough identities
\begin{eqnarray}
\beta  \; (-i \alpha^{0})  = \Gamma^{0} \; ,\;\; \beta \; \alpha^{j} = \gamma^{j}\;,
\qquad  \mbox{where}\qquad  \beta = \left | \begin{array}{rrrr}
1 & 0  &  0  & 0  \\
0 & -i  & 0  & 0  \\
0 & 0  &  -i  & 0  \\
0 & 0  &  0  & -i
\end{array}  \right | \;  .
\label{a17}
\end{eqnarray}

Esposito's representation of the Maxwell equation at any 4-vector $u^{\alpha}$ can be easily related to the matrix
equation of Riemann -- Silberstein -- Majorana -- Oppenheimer in the form  (\ref{a16}):
\begin{eqnarray}
(-i \alpha^{0} \partial_{0} + \alpha^{j}\partial_{j}  )   \Psi   = J \;  , \qquad \Longrightarrow
\label{a18}
\\[2mm]
(-i \alpha^{0} \partial_{0} + \alpha^{j}\partial_{j}  )   U^{-1}  \; ( U \Psi  ) = J \;  ,
\nonumber
\\
-i\alpha^{0} \; U^{-1} = \beta \;\Gamma^{0} \;, \qquad  \alpha^{j}\; U^{-1} = \beta \;\Gamma^{j} \; , \qquad
 U \Psi  = \Phi \;,
\nonumber
\\
\beta \; ( \Gamma^{0} \partial_{0}  + \Gamma^{j} \partial_{j} ) \; \Phi = J \;  ,
\qquad \beta^{-1}J = \epsilon_{0}^{-1} (j^{a})\; ,
\nonumber
\\[2mm]
( \Gamma^{0} \partial_{0}  + \Gamma^{j} \partial_{j} )  \Phi =  \epsilon_{0}^{-1} \; j  \; .
\label{a19}
\end{eqnarray}

Eq.  (\ref{a19})   is  a  matrix representation of the Maxwell equations in Esposito's  form
 \begin{eqnarray}
\partial_{\alpha} \; [ \;  \delta^{\alpha}_{\gamma} u^{\beta}   -
  \delta^{\beta}_{\gamma} u^{\alpha}   +
i \epsilon^{\alpha \beta \rho \sigma}  g_{\rho \gamma}  u_{\sigma}\; ]\; \Phi^{\gamma} =
\epsilon_{0}^{-1} j^{\beta} \; .
\label{a20}
\end{eqnarray}

Evidently,  eqs. (\ref{a18}) and (\ref{a19}) are equivalent to each other. There is no ground to consider the form
(\ref{a19}) -- (\ref{a20}) obtained through the trivial use of identity $I = U^{-1}(u) U(u)$ as having
certain  especially profound sense. Our point of view  contrasts with the claim
by Ivezi\'c
\cite{2001-Ivezic(1)}-\cite{2002-Ivezic}-\cite{2002-Ivezic(2)}-\cite{2002-Ivezic(3)}-\cite{2003-Ivezic}-
\cite{2005-Ivezic(1)}-\cite{2005-Ivezic(2)}-\cite{2005-Ivezic(3)}-\cite{2006-Ivezic} that eq. (\ref{a20})
has a status of a true Maxwell equation in  a moving reference frame (at this $u^{\alpha}$ is identified with 4-velocity).

\vspace{5mm}

\section{Acknowledgement}

Authors are grateful to  Kurochkin Ya.A.  and Tolkachev E.A. for discussion
and advice. This  work was    supported  by the Fund for
Basic Research of Belarus, grant F07-314.

\end{document}